# Spherical Distance Metrics Applied to Protein Structure Classification*

**James DeFelice[1] and Vicente M. Reyes, Ph.D.[2], [3]**

(*, M. S.  Thesis; [1], M. S. student; [2], thesis advisor)
[3], E-mail:  **vmrsbi.RIT.biology@gmail.com**

Submitted in partial fulfillment of the requirements for the **Master of Science** degree
in Bioinformatics at the Rochester Institute of Technology

**James DeFelice**

Dept. of Biological Sciences, School of Life Sciences
Rochester Institute of Technology
One Lomb Memorial Drive, Rochester, NY 14623

May 2011

**Rochester Institute of Technology**
**RIT Scholar Works**

5-18-2011

# Spherical distance metrics applied to protein structure classification

James DeFelice

# Spherical Distance Metrics Applied To Protein Structure Classification

*James DeFelice*

Master of Science in Bioinformatics

Department of Biological Sciences

School of Medical and Biological Sciences

College of Science

Rochester Institute of Technology

Approved on May 18[th], 2011

*Dr. Vicente Reyes*

*Dr. James Halavin*

*Dr. Michael Osier*

# Abstract:

Observed protein structures usually represent energetically favorable conformations. While not all observed structures are necessarily functional, it is generally agreed that protein structure is closely related to protein function. Given a collection of proteins sharing a common global structure, variations in their local structures at specific, critical locations may result in different biological functions. Structural relationships among proteins are important in the study of the evolution of proteins as well as in drug design and development.

Analysis of geometrical 3D protein structure has been shown to be effective with respect to classifying proteins. Prior work has shown that the Double Centroid Reduced Representation (DCRR) model is a useful geometric representation for protein structure with respect to visual models, reducing the quantity of modeled information for each amino acid, yet retaining the most important geometrical and chemical features of each: the centroids of the backbone and of the side-chain. DCRR has not yet been applied in the calculation of geometric structural similarity.

Meanwhile, multi-dimensional indexing (MDI) of protein structure combines protein structural analysis with distance metrics to facilitate structural similarity queries and is also used for clustering protein structures into related groups. In this respect, the combination of geometric models with MDI has been shown to be effective. Prior work, notably Distance and Density based Protein Indexing (DDPIn), applies MDI to protein models based on the geometry of the Cα backbone. DDPIn's distance metrics are based on radial and density functions that incorporate spherical-based metrics, and the indices are built from metric tree (M-tree) structures.

This work combines DCRR with DDPIn for the development of new DCRR centroid-based metrics: spherical binning distance and inter-centroid spherical distance. The use of DCRR models will provide additional significant structural information via the inclusion of side-chain centroids. Additionally, the newly developed distancemetric functions combined with DCRR and M-tree indexing attempt to improve upon the performance of prior work (DDPIn), given the same data set, with respect to both individual k-nearest neighbor (kNN) search queries as well as clustering all proteins in the index.

# I.  Background

In a biological system a protein's structure, at the atomic level, determines its function. Therefore it is useful to compare proteins at a structural level to gain insight into the role that a protein plays within a particular biological setting. Proteins are often classified and grouped according to specific structural properties, sometimes in an automated fashion and at other times via manual curation.

Protein structure is often described either in linear (primary sequence) or more complex terms (secondary, tertiary, quaternary, 3D). Although it can be useful in determining protein function, protein sequence can vary widely among structurally similar proteins, and even more so for distantly related proteins with large structure variations. Higher levels of protein structure are accessible in three-dimensional space, and in such space the relative proximity among various protein components becomes important for understanding protein function. This work focuses on protein 3D structure.

Various methods have been proposed for navigating protein structure in 3D space, most of which tend to focus on the location, and relative positions of the $C\alpha$ and $C\beta$ atoms (see Figure 1). For example, Sequential Structure Alignment Program (SSAP) [1] operates using $C\beta$ distance vectors to compute protein alignments. Another popular method, Combinatorial Extension (CE) [2], groups protein residues into aligned fragment pairs (AFPs) and for each AFP substructure calculates a set of distances between the $C\alpha$ atoms. The substructures are compared and extended (based on these distance distributions) in a manner that minimizes one of several distance metrics. The authors of CE note that the method lacks support for "... non-topological similarities, where the order of polypeptide fragments in the structure does not follow their order in the sequence" [2]. Other tools operate on an all-atom representation, meaning that every atom of the protein is taken into consideration for analysis. As opposed to all-atom-based methods, tools that use reduced model representations have been successful in several aspects of protein analysis, modeling, and visualization.

One such reduced representation is DCRR [3], which is constructed from the backbone and side-chain atom coordinates for each residue. A DCRR model represents a protein structure using centroids, where the centroids are modeled from the backbone ($\Psi bb$) and side-chain ($\Psi sc$) atoms for every amino acid. Given that the number of elements in a DCRR model is greater than that of a purely $C\alpha$ (or $C\beta$) model, more information is available for structural analysis. The fact that less information is available in DCRR than that of an all-atom model should not pose a problem for analysis: protein structure analysis tools have been relatively successful in obtaining good results with less data than DCRR provides. DCRR should provide *just enough information* to allow protein structure analysis tools to produce even better results than when those tools operate on purely Cα or Cβ models.

## 1.1. Viewpoints

Protein structure comparison may be likened to other types of general shape comparison. Recent work points to a *single viewpoint perspective*, wherein the structure is centered at the origin in $\Re^3$ according to its center-of-mass [4] [5]. Such algorithms are typically looking to compare global shape structure. Some of these algorithms attempt to normalize the rotation of the shape around its center of mass, because the extracted features used for comparison are sensitive to initial rotation [6] [7]. This normalization process can be error-prone, and prior experiments have shown that similar objects, which are relatively star-shaped, can be difficult to normalize in a consistent manner.

Two or more shapes are sometimes compared simultaneously [8] [9], and many algorithms exist which attempt to minimize the Root Mean Square Deviation (RMSD) of the distance between analogous points among several structures. In doing so, the structures are aligned and overlapped in $\Re^3$. Such algorithms are useful for proteins which share highly similar, common core shape patterns but show weakness when comparing more distantly related shape structures.

Another class of approaches work with a *multiple viewpoint perspective*. SSAP is the basis of a family of very successful protein structure analysis tools: it uses a combination of Dynamic Programming (DP) and Cβ atom position to compute features for each residue in the protein. In the SSAP model, each reside provides a viewpoint for feature extraction and the features are rotationally invariant. Variants of SSAP have been implemented that optimize search speed as well as perform multiple structure alignments.

Prior, in-house experimentation has shown that there is a significant amount of information in the radial coordinate of protein shape data when the protein is centered at the origin according to its geometric center and modeled in spherical space. Variants of both single- and multiple-viewpoint algorithms make use of radial information in different ways. One common pattern seems to be the construction of concentric spheres around a viewpoint, and then sampling the space between the spheres (or in the case of the inner-most sphere, sampling the entire sphere). Such samples then form the basis of a data set from which features are extracted.

## 1.2. Viewpoint Semantics

DDPIn [10] extends the multiple viewpoint concept by extracting several feature sets simultaneously. The features form the basis of a metric search space that is navigable via kNN-based queries (a good reference for kNN and related classification techniques is [11]). The algorithms used to extract features in DDPIn are referred to as viewpoint semantics. A combination of variable and fixed length radial semantics (sRad and sDens) provided the best results.

In DDPIn, the set of all the feature variables $\{x_i\}_{i=0}^{n}$ for each of the n concentric spheres centered at a particular viewpoint ($\Psi x$) form the feature vector for that centroid. Given:

$$rad(x_i) : \text{distance from } \Psi_x \text{ to edge of } i^{th} \text{ sphere}$$

$$width(x_i) : \begin{cases} rad(x_i) - rad(x_{i-1}) & i > 0 \\ rad(x_0) & i = 0 \end{cases}$$

$$dens(x_i) : \text{sum of residues in } i^{th} \text{ ring}$$

Variable length radius semantics (sRad) are defined as:

$$\forall i, j : width(x_i) = width(x_j) = w$$

$$\forall i : P_i^x = dens(x_i)$$

where $p\Psi c$ is some percentage of all the centroids of class c in the protein structure and $P^x$ is the feature vector extracted using the sRad semantic.

Fixed length radius semantics (sDens) are defined as:

$$\forall i, j : dens(x_i) = dens(x_j) = p_{\Psi_c}$$

$$\forall i, j : P_i^x = rad(x_i)$$

where ω is some fixed radius width and $P^x$ is the feature vector extracted using the sDens semantic.

For example, given the protein strand and viewpoint centroid x as illustrated in Figure 3(A), the sDens semantic would generate feature vector $P^x$ = (1, 9, 3) for i ∈ [0..2].

## 1.3. Distance

The authors of DDPIn utilized $L^2$ and weighted-$L^2$ metrics to compute the distance between feature sets. Weights for the weighted-$L^2$ metrics were multiplied by each partial difference in normal $L^2$ calculation. The weights that gave greater emphasis to closer regions (vs. those more distant) were more successful.

Given weight vector w and vector δ that contains the squares of the differences between equal length vectors $P^x$ and $P^y$:

$$L_{weighted}^{2}\left(x,y\right) = \sqrt{w^T \cdot \delta} \qquad (1)$$

Given D, a number of dimensions less than or equal to $|P^x|$, vector w is calculated as follows:

$$w_i = log\left(D - i + 2\right), i \in 1...D \qquad (2)$$

DDPIn uses the following scoring function when determining the structural similarity of a single query protein q against a previously indexed protein p:

$$s(p,q) = \sum_{i=0}^{|\Psi_q|-1} \sum_{j=0}^{|\Psi_p|-1} \begin{cases} k - \log(z), & z \in 1...k, \\ & \text{if } P^{\hat{p}} \in NN(P^{\hat{q}}), \\ \\ 0, & \text{otherwise} \end{cases}$$

where $|\Psi_p|$ and $|\Psi_q|$ indicate the number of centroids being evaluated for proteins p and q, $\hat{p} = \Psi_p(j)$ and $\hat{q} = \Psi_q(i)$ indicate the centroids at indices j and i for proteins p and q, NN(x) is the set of nearest-neighbors for viewpoint $P^x$, k is the size of the set and z is the relative index of $P^{\hat{p}}$ in the set. Higher s scores indicate a more likely match between proteins p and q.

Rong Li et al. explore various distance metrics for comparing data sets and have settled on the Wasserstein metric as the most appropriate [12]. According to their work, Wasserstein distance may be computed as the *average pairwise distance* when the sizes of the data sets are the same and the sets consist of scalar data points:

$$d(X, Y) = \frac{1}{n} \sum_{i=1}^{n} d(x_{(i)}, y_{(i)}) \qquad (3)$$

where X and Y are data sets, ($\{x(i)\}_{i=0}^{n}$, $\{y(i)\}_{i=0}^{n}$) are ranked values of (X, Y), and d($\cdot$, $\cdot$) is some distance function. When X, Y are sets of vectors the *average minimum pairwise distance* can be used to compute their difference:

$$d(X, Y) = \min_{I \in \mathcal{I}} \frac{1}{n} \sum_{i=1}^{n} d(x_{(i)}, y_{(i)}) \qquad (4)$$

where $\mathcal{I}$ is the set of all possible permutations of Y, $I \in \mathcal{I}$ is a single permutation of Y yielding $\{y(i)\}_{i=1}^{n}$, and d($\cdot$, $\cdot$) is some distance function ($L^p$ norm is recommended in [12]). Again, X and Y must be equivalent.

In the case that X and Y are not equivalent, the more general form of Wasserstein distance ($L^p$) may be applied (for p = 1, 2):

$$l_1(F, G) = \int_0^1 |F^{-1}(t) - G^{-1}(t)| dt \qquad (5)$$

$$l_2^2(F, G) = \int_0^1 \left[ F^{-1}(t) - G^{-1}(t) \right]^2 dt \qquad (6)$$

where F and G are cumulative distribution function (CDF)s of independent random variables. $F^{-1}$ and $G^{-1}$ are quantile functions:

$$F(x) = Pr(X \le x) = f \qquad (7)$$

$$F^{-1}(f) = \inf\{x \in \mathfrak{R} : f \le F(x)\} \qquad (8)$$

for any probability f, 0 < f < 1.

## 1.4. M-tree

Metric access methods (MAMs) permit efficient exploration of a metric space through the use of index structures and distance metric functions, allowing large parts of the space to be excluded during search operations. MAMs have been successfully applied in several contexts and remain relevant to those of a biological nature [13] [14]. Notably, the Mtree [15] structure can provide a reasonably good method of clustering multidimensional data sets using appropriate distance metrics. The user of the M-tree is typically responsible for providing the distance metric algorithm. Nearest neighbor and range queries are the most useful query types satisfied by M-tree structures.

M-tree nodes can be classified as either being leaf or nonleaf. Non-leaf nodes contain routing information that allows search algorithms to limit the scope of their processing to only the most relevant nodes. Leaf nodes contain the information in the data set. As data objects are added to the M-tree, the distance algorithm is repeatedly applied, along with non-leaf routing information, to organize the new data objects appropriately (adds each data object to the leaf with the nearest neighbors). It has been previously reported [16] that overlap between leaf nodes can cause inefficiencies in the search space, however the optimization of leaf node overlap will not be addressed here.

The primary interest of this work is the multidimensional indexing and clustering capabilities of the M-tree. In [10] the authors use M-trees to create indices of protein structure features and show that M-tree performance can be acceptable for protein structure queries based on multiple viewpoint semantics. At least two implementations of M-tree are currently freely available [17] [18].

Related to the M-tree is the onion-tree structure [19]. The onion-tree is an in-memory (vs. on-disk) MAM index designed to minimize the number of distance calculations required for kNN and range-based queries. All onion nodes contain, at most, two data objects along with a fixed or variable number of other child nodes (different configurations can use fixed or variable numbers of child nodes). These two objects, or pivots, are used to slice a search space into non-overlapping searchable regions. The non-overlapping regions are particularly useful to the search algorithm, which optimizes the order in which such regions are visited in order to eliminate unnecessary distance calculations. An implementation of the onion-tree source code is currently available online [20].

## 1.5. Partitioning Algorithms

This work will utilize a new method of partitioning, or *binning* the surface of a sphere. Two approaches were initially investigated:

1. Partition the surface of a sphere by slicing at regular intervals of φ and θ, and also;

2. Center the sphere at the origin in 3D Euclidean space, slice along the x, y and z axis to create eight wedges (the surface of each wedge is spherical triangle), and then recursively subdivide the resulting spherical triangles by connecting the midpoints of each arc.

The first approach, *polar binning*, is suggested by Reyes and, admittedly, it suffers from non-uniform bin sizes [21]. As a work-around, the author suggests normalizing bin values by computing bin density, dividing the initial bin value by the area of the bin. However, the gross differences, especially at the poles, in bin size result in a very non-uniform sampling of the spherical surface and could therefore introduce artifacts into differencing algorithms, even when normalized by bin area.

The second approach resulted in non-uniform spherical triangles. Consider the unit sphere, bisected along each plane in $\Re^3$ to form eight initial wedges. The spherical triangles at the surface of the initial wedges are uniform, the sides of each being 1.5708 radians. After one iteration of subdivision, each of the original wedges contains four smaller wedges. The smaller wedges are not consistently uniform, three of them are isosceles (the smallest wedges in each corner of the parent wedge) with sides of (0.7854, 0.7854, 1.0472) radians, and the central wedge is significantly larger than the prior three with sides of (1.0472,1.0472,1.0472) radians.

Next, the investigation focused on polyhedrons, the idea being that the vertices of polyhedrons could be projected to the surface of a sphere to create a mesh. Polyhedrons with equilateral polygonal faces are convenient to work with because each face may be subdivided into child faces that retain the equilateral properties of their parent. Though experimentation, equilateral quadrilateral polygon faces showed the highest degree of uniformity when recursively projected and subdivided.

Both the icosahedron (equilateral triangle faces) and the rhombic triacontahedron (rhombic faces) were examined in detail. The best way to subdivide an equilateral triangle into four smaller equilateral triangles is to connect the midpoints of each of the sides of the sides of the triangle. An easily observed problem with this approach is that upon recursive projection and subdivision, the resulting triangular faces toward the center of the original polyhedron face become much larger in area than those triangular faces near the corners of the original face. Equilateral quadrilaterals suffer less from this phenomenon due to the fact that the center portion of the face is shared more equally by the child faces derived from the projection/subdivision process, so that when projected to a spherical surface, the area of the child faces is much more uniform.

Thus, it is the later of the polyhedron forms that seems to give the most consistent, uniform bin sizes when subjected to the following binning process:

1. Begin with a rhombic triacontahedron (convex polyhedron, 30 faces, each face a rhombus);

2. surround the polyhedron with a sphere;

3. recursively apply the following steps:

    (a) project all vertices of the polyhedral lattice to the surface of the surrounding sphere, then;

    (b) bisect each face of the polyhedron twice to create four nearly identical, nearly rhombic child faces from each original face.

Three or more levels of recursion will result in at least thousands of bins, the surface areas of each bin being very similar to each other. The resulting bins form a spherical mesh that is easily scaled by manipulating the radial coefficient of each vertex (assume spherical coordinates).

Of further interest to readers may be Kulikowski's Theorem [22], which deals with identifying an arbitrary number of lattice points along the surface of the sphere.

## 1.6. Quad Trees

The process of reducing a spherical surface to trixels, as described by Fekete, leverages an icosahedron for the initial triangulation of the surface [23]. Instead of trixels, this work leverages the rhombic faces described in the prior section. Thus, each rhombic face becomes a node in the quad tree, capable of holding exactly four child nodes if and when it is sub-divided. The polyhedral structure described above, when combined with quad tree mechanics, offers several opportunities for optimization.

First, it is important to map the geometric center of the polyhedron and the geometric center of the protein to the same origin. Next, each centroid is modeled as a ray, originating at the geometric origin and passing through the centroid. Every projected centroid ray will intersect a single rhombic face, or quad-tile, regardless of the number of times each original quad-tile has been sub-divided. However, as the number of rhombic subdivisions increases, the number of tiles that could possibly be evaluated grows exponentially This becomes a problem since ray/quad-tile intersection is an expensive operation.

The first optimization addresses this problem. The basic premise is to, for each ray, search *only* the child nodes of the parent node that is intersected by the ray. This optimization is only effective if the original quad-tiles are sub-divided.

However, it is possible to extend the optimization such that the top-level quad-tiles are grouped into *super-root* nodes. This super-root optimization can be applied in cases even where the original quad-tiles have not been subdivided. When building the rhombic polyhedron, it is possible to overlay a tetrahedron in the same geometrical space such that the tetrahedron shares its vertices with those of the rhombic polyhedron. The triangular faces of the overlaid tetrahedron will overlap a subset of the faces of the rhombic polyhedron. Each tetrahedral face thus defines boundaries for grouping the rhombic faces into super-roots: If a projected ray intersects a particular tetrahedral face (super-root) then it also must intersect one of the overlapping rhombic faces. This allows the algorithm to limit the number of root quad-tiles that are checked for ray intersection by only checking those tiles that overlap with the intersected super-root.

Another optimization is related to the process of quantifying the number of intersections per quad-tile. The number of intersections per quad-tile is important for the calculation of ray density per quad-tile. Every time a ray intersects a leaf node of the quad-tree, the density for that leaf node increases by one. Every time a node (parent or child) density value is increased, the density of that node's parent is also increased. By tracking node densities in this manner it becomes possible to optimize algorithms that use node density by pruning parent nodes with density values of zero, saving the cost of visiting all the empty, descendant child nodes.

## 1.7. Quaternions

A quaternion exists in four-dimensional space H, three of which are imaginary. Quaternions may be written as $w + ix + jy + kz$ where $w$ is real, and $i$, $j$, $k$ imaginary coefficients of $x$, $y$, $z$. The primary use of quaternions in this work is for simplified rotation in $\Re^3$ about an arbitrary axis.

Given a *normalized* vector as an axis, $\vec{n}$ and an angle of rotation θ, quaternion q and its conjugate q′ may constructed as:

$$q = \cos\left(\frac{\theta}{2}\right) + i\left(n_x \sin\left(\frac{\theta}{2}\right)\right) + j\left(n_y \sin\left(\frac{\theta}{2}\right)\right)$$
$$+ k\left(n_z \sin\left(\frac{\theta}{2}\right)\right)$$
$$q' = \cos\left(\frac{\theta}{2}\right) - i\left(n_x \sin\left(\frac{\theta}{2}\right)\right) - j\left(n_y \sin\left(\frac{\theta}{2}\right)\right)$$
$$- k\left(n_z \sin\left(\frac{\theta}{2}\right)\right)$$

The process for rotating a point p (in Cartesian space) around axis $\vec{n}$ for θ radians (to yield $p_1$) is as follows:

1. Convert p into quaternion form ~p: $\tilde{p} = 0 + ip_x + jp_y + kp_z$

2. Perform the rotation : $\tilde{p}_1 = q \cdot \tilde{p} \cdot q'$

3. Convert ~p1 into Euclidean form : $p_1 = [\tilde{p}_{1x}\ \tilde{p}_{1y}\ \tilde{p}_{1z}]^T$

## 2. Methods

This research intends to show that DCRR models can be useful for protein structure comparison and classification when such models are paired with appropriate structural distance metric functions and classification methods. The use of the additional information provided by DCRR models (vs. pure Cα or Cβ) will attempt to provide results as good as, or better than existing models. The "gold standard" reference used to assess classification performance is Structural Classification Of Proteins (SCOP) [24], a manually curated collection of structural classifications.

To leverage the additional information provided by DCRR, several new protein structure comparison methods are developed. The focus of these methods is to calculate distance metrics in order to quantitatively describe the differences between two protein structures. These methods are pluggable components in the context of a protein structure analysis pipeline. Some of the new methods are used for indexing protein structures, while others are used for alignment in the voting process.

## 2.1. Protein Data Set

This work uses a protein data set constructed in a manner similar to that of prior experiments in order to facilitate a comparison of relative strength/weakness, with regard to classification power, between this and prior implementations. The construction of the list is formally described by Camoglu et al. [25] and is repeated here for ease of reference:

   1. Build a working set W of proteins belonging to one of the following SCOP classes: all-α, all-β, α+β, α/β.

   2. Choose SCOP superfamilies S that have at least ten proteins in W ($S_1$, $S_2$..$S_x$).

   3. Build a working set $W_1$ by sampling ten proteins from the set of each superfamily ($S_n$).

   4. Then build a query set $Q_1$ by picking a random protein from each $S_n$ not already in $W_1$.

   Query proteins are not chosen at random, but instead are reused from prior work [25]. The list of proteins in $Q_1$ appears in Table 6. The proteins of $W_1$ are chosen at random and little data cleaning is applied in the process of selection - only proteins without any computable centroids are eliminated. SCOP version 1.75 is used as the gold standard of protein structure classification; all classification results are compared to this version.

## 2.2. Semantics

As in DDPIn this experiment will leverage multiple viewpoints along the protein chain to derive indexed feature sets for the purposes of structural comparison [10]. A viewpoint semantic is a method for extracting a feature from a particular viewpoint. Each semantic requires at least two inputs: the origin of the viewpoint and the locations of the centroids that are to be sampled.

   Since DCRR models provide more interesting locations along the chain to sample from (both the Ψbb and Ψsc centroids), this experiment will vary both the origin centroid and the sampled centroids in order to generate different indices. The data sets (origin centroid, sampled centroids) are generated from the following permutations of the DCRR model (Ψbb, Ψbb), (Ψsc, Ψsc) and (Ψbb + Ψsc, Ψbb + Ψsc).

   DDPIn explored the use of various semantics for feature extraction, the most useful being sRad and sDens. This experiment reuses the sRad semantic, in part to validate the integrity of the experimental setup and also to provide insight into the information gained by introducing additional side-chain centroids in the viewpoint model. Like DDPIn, a weighted $L^2$ distance metric is applied to calculate the similarity between viewpoint features.

In addition to the sRad semantic above, this experiment introduces several new semantics that attempt to produce results as good as, or better than the best DDPIn semantics. Several semantics may be derived by projecting the sampled centroids to the surface of the nearest outer concentric sphere and analyzing the resulting spherical distributions in different ways. Several forms of variable-length radial analyses are applied. Feature vectors are constructed from:

- sSumDist: the sum of the spherical distances between all projected points, and;

- sNormDist: a normalized spherical distance between all projected points, using an arithmetic or geometric mean.

Assuming spherical coordinates $(\rho, \phi, \theta)$, the spherical distance between two points $(s, f)$ where $\rho_s = \rho_f$ may be represented as $\rho\Delta\hat{\sigma}(s, f)$, where:

$$\Delta\hat{\sigma} = \arccos\left(\sin\phi_s \sin\phi_f + \cos\phi_s \cos\phi_f \cos\Delta\theta\right) \quad (9)$$

If s and f are instead represented as vectors in 3D Euclidean space, then the spherical distance may be computed as a function of their dot product:

$$\rho\Delta\hat{c}(s, f) = \rho \arccos\left(\frac{s \cdot f}{\|s\| \|f\|}\right) \quad (10)$$

For some fixed-length radial semantics the $\rho$ factor may be left out completely (in some cases the protein structure is normed to fit inside the $S^2$ sphere). Consider $\Delta\hat{\sigma}$, a set of pairwise inter-centroid spherical distances:

$$\forall \Delta\hat{\sigma} - \begin{bmatrix} 0 & \Delta\hat{\sigma}_{1,2}, & \cdots & \Delta\hat{\sigma}_{1,k-1} & \Delta\hat{\sigma}_{1,k} \\ & 0 & \cdots & \Delta\hat{\sigma}_{2,k-1} & \Delta\hat{\sigma}_{2,k} \\ & & \ddots & & \vdots \\ & & & 0 & \Delta\hat{\sigma}_{k-1,k} \\ & & & & 0 \end{bmatrix} \quad (11)$$

where $X_i^x$ is the set of projected centroids for sphere i centered at $\Psi x$, K is the size of $X_i^x$, j, k $\in$ [1,K], $\Delta\hat{\sigma}j,k \sim \Delta\hat{\sigma}(s_j, f_k) : s_j, f_k \in X_i^x$. sSumDist may be calculated as follows:

$$\forall i : P_i^x = \rho \sum_{j=1}^{K-1} \sum_{k=j+1}^{K} \Delta\hat{\sigma}_{j,k} \qquad (12)$$

sNormDist, by extension, may be written as:

$$\forall i : P_i^x = \frac{\rho}{K(K-1)} \sum_{j=1}^{K-1} \sum_{k=j+1}^{K} \Delta\hat{\sigma}_{j,k} \qquad (13)$$

Similarity of features sSumDist and sNormDist are computed using a weighted $L^2$ metric. Features generated from sDistDist are compared according to Wasserstein distance.

## 2.3. Indexing

Database construction is accomplished though a mostly reusable software pipeline that begins with the input of PDB [26] files and result in the generation of multiple indices. The construction has clearly defined extension points that allows new semantics and distance functions to be easily integrated. Protein features are extracted from both $Q_1$ and $W_1$ and stored as compressed objects. The semantics, distance functions, and segmentations listed in Table 1 are used for feature extraction.

Segmentation for sRad is computed differently than for the other two. For sRad a segmentation of 14 means that each concentric ring segment will contain 1/14 of the proteins overall density. For sSumDist and sNormDist a segmentation of 14 is calculated according to the following rules:

1. Determine $d\Psi x$, the distance to the centroid $\Psi x$, which is the furthest from the geometrical center of the protein.

2. Determine $d_{max} = 2 * d\Psi x$, the maximum possible distance between $\Psi x$ and any other centroid of the protein.

3. Determine $d_{seg} = d_{max}/14$, the distance between the perimeter of each concentric ring (segment).

At run-time, the compressed feature objects of $W_1$ are read into an in-memory MAM-based index, constituting the database. Both M-tree and onion-tree structures are used as these databases. Database index searches use the kNN algorithm, identifying the most closely related proteins. An index is queried only with features built from the same semantic as were used to construct the index. Queries are executed that use both full and partial sets of feature dimensions. For example, a feature vector of length 14 may be truncated to a length of 11 for a given query.

Similarly, the database search process is defined by a software pipeline (partially automated) that establishes clearly defined integration points for new, or different combinations of feature sets. The product of the search pipeline is a set of the most closely related proteins and their distance scores, as well as several tables and graphs for use in understanding the quality of the results. Figure 4 illustrates both the database construction and query processes.

## 2.4. Classification

Three classification methods are evaluated, where classes correspond to SCOP superfamilies. Each classification function $\omega_m(P)$ estimates the correct class for protein P.

• $\omega_1(P)$ : choose the class of the highest ranked neighbor of the kNN search.

• $\omega_2(P)$ : from within the classes of the five highest ranked neighbors of the kNN search, choose the most frequently occurring class; if all classes occur with equal probability, then the highest ranked class is chosen.

• $\omega_3(P)$ : from within the classes of the five highest ranked neighbors of the kNN search, choose the class with the highest combined score.

Precision of the classifiers is calculated somewhat differently than is typical for search algorithms. kNN-based queries will only return a maximum of k results. Precision @ N is a metric used in such cases to calculate the precision within the top N results of the query. This work adjusts the above metric such that N becomes $N_{kf}$ a percentile of the overall query results (for example, the top ten percent). Precision @ $N_{kf}$ is calculated, where $N_{kf} = \max(1, k/f)$ and $f \in (1, 2, 3, 4, 5, 10, 15, 100)$, which accounts for situations when $f < k$ by forcing $N_{kf} = 1$. Average precision @ N for this experiment is calculated as:

$$AvgPr(k) = \frac{\sum\limits_{q=1}^{|Q_1|} Pr_q(N_{kf})}{|Q_1|} \qquad (14)$$

$$Pr_q(N_{kf}) = \frac{\sum\limits_{r=1}^{|R_q|} rel(r_i)}{N_{kf}}, r_i \in R_q \qquad (15)$$

where $R_q$ is the set of ranked results of a kNN query for query protein q.

## 2.5. Voting

An integral component of the search pipeline is a voting mechanism (the authors of DDPIn called this "healing") by which the search results from two index searches are compared. Classifications that disagree are passed to a voting process. Classifications that agree are evaluated for correctness. The voting process evaluates the query protein in the context of the neighborhood of the kNN proteins and classifies the query protein according to its closest match. Closeness is computed according to an alignment-based distance function.

The voting criteria in DDPIn was generated by a Smith-Waterman DP matrix, based on the primary structure of the query protein and the primary structures of the proteins in the nearest neighborhood, for each index. In this experiment, voting is implemented by spherical alignment algorithms and the effectiveness of each algorithm is compared. Spherical alignment applies partitioning procedures that take advantage of the representation of proteins in spherical space [21]. For a discussion of partitioning procedures refer to section 1.5.

The alignment of two proteins is performed by the following steps:

1. Surround two proteins A and B each with their own partitioned spheres;

2. project the DCRR centroids of each protein to its surrounding sphere;

3. calculate structural distance by comparing the partition sets of the two spheres;

4. systematically rotate one or both of proteins A and B, and finally;

5. repeat the calculation and rotation steps until all rotations have been assigned distance metrics.

The rotation that yields the least distance between partition sets is the best alignment of the two proteins. Once the voting process has gathered the results of all the protein alignments for a given query protein, the database protein that is most closely aligned with the query protein determines the estimated superfamily.

## 2.5.1. Rotation

Rotation is performed on two protein structures A and B, each of which has an identified origin at its geometric centroid, that have been translated such that their origins coincide. Prior to the actual rotation of the proteins, compute the rotational coefficients:

1. Compute a set of normalized vectors _ distributed uniformly in $\Re^3$, where $\forall a \in [1...n]$, $\lambda a$ is rooted at the origin. One way to accomplish this would be to create a spherical mesh, centered at the origin, and use all n normalized vectors of the mesh vertices.

2. Compute a set of angles $\Theta$, where $\theta_t \in [0, 360]$ in multiples of $\Delta\hat{\theta}$, t is the number of elements in $\Theta$. For example, $\Delta\hat{\theta} = \pi/30$ indicates increments of six degrees.

3. Compute a set of quaternions (see section 1.7) $\Upsilon$ such that $v_{at}$ represents the quaternion for rotation about $\lambda_a$ for all $\theta_t$ degrees.

4. Compute a set of conjugate quaternions $\Upsilon'$.

Keep the rotation of protein structure A fixed and iteratively rotate protein structure B. Given function $\Gamma(P)$ that generates quaternions for all points p in Euclidean structure P, and function $\Gamma^{-1}(Q)$ that generates Euclidean points for all quaternions q in quaternion structure Q:

$$\tilde{B} = \Gamma(B) \qquad (16)$$

$$\forall \{v_{at} \in \Upsilon, v'_{at} \in \Upsilon'\} : \tilde{B}_{at} = v_{at} \cdot \tilde{B} \cdot v'_{at} \qquad (17)$$

$$B_{1at} = \Gamma^{-1}(\tilde{B}_{at}) \qquad (18)$$

$$d_{at}\{\in D\} = d(A, B_{1at}) \qquad (19)$$

where $d(\cdot, \cdot)$ is a distance metric function for two protein structures. The structural distance between A and B is defined as the minimum value in D.

## 2.5.2. Comparing Partition Sets

Structural distance may be computed by different methods, not altogether much different than prior methods suggested for viewpoint semantics.

Method 1 (global, $dG(\cdot, \cdot)$):

1. Form a single surrounding sphere;

2. project each centroid to the surrounding, partitioned sphere;

3. assign each partition a value (e.g. the number of centroids that occupy it), and;

4. compare partition distributions (using a Wasserstein or some other, related metric).

Method 2 (concentric, $dL(\cdot, \cdot)$):

1. Form several concentric spheres, isolating the DCRR centroids at predetermined radial intervals;

2. project each centroid to its closest surrounding, partitioned sphere;

3. for each concentric sphere, assign partition values - each protein will have a set of distributions, one for each concentric sphere;

4. cast the set of data for each protein as a multivariate feature vector, and;

5. apply a multivariate feature comparison metric to determine similarity.

Each partition value is the sum of the number of projected centroids in that particular partition. Determining the partition counts is expensive and relies upon a ray/quad-tile intersection algorithm. The number of intersection computations is reduced through the use of quad trees. For additional details see section 1.6.

## 2.6. Platform

Feature extraction and database searches both heavily depend upon CPU resources. All feature extraction, database searches, and voting alignment algorithms are implemented in Java and executed on modern 64-bit Java virtual machines (JVMs) (IcedTea 1.7.5 b17 and Oracle JDK 1.6.0 b21-b25) running on the Linux operating system. Many different compute nodes are used to generate the results (see Table 2).

In order to minimize data preparation time, feature object builds are distributed across available compute nodes.

## 3. Results

Sample build times for feature objects constructed from $W_1$ appear in Table 3. Builds were generated on compute nodes with up to eight cores using up to four cores per build process.

The size of the compressed W1 feature object sets on disk varies between 21.5MB and 185.2MB. Compressibility among W1 feature object sets varies, as can be seen in Figure 5. Compression factors are calculated as ($\text{size}_{KB}/(|\Psi x| \times \text{segments})$), where $|\Psi x|$ is the number of centroids in the feature set. Values of $|\Psi \text{all}|$, $|\Psi \text{bb}|$ and $|\Psi \text{sc}|$ are 1163769, 593566 and 570203 respectively.

## 3.1. Single Semantic Classification

With respect to the different centroid sets, $\Psi \text{bb}$ always gave the highest scoring and most consistent results. $\text{sRad}_{k=1}^{14/14}$, $\text{sRad}_{k=1}^{16/16}$ and $\text{sRad}_{k=1}^{17/18}$ were tied as the highest scoring semantics. The next best centroid set was $\Psi \text{all}$ (ranking 2nd place in a six-way tie) for classifier $\omega_1$ and semantic $\text{sSumDist}_{k=1}^{14/14}$. Regardless of the classifier function, the sRad semantic consistently generated the highest scoring classifications when compared to sSumDist and sNormDist. Semantic $\text{sSumDist}_{k=1}^{16/16}$ yielded the most consistent scoring percentage of 70 percent across all three classifiers. Of all the classifiers, $\omega_1$ generated the highest scoring classifications (see Table 4). A single "best set" of classification scores for each of the semantics is shown in Figure 6.

Precision was the most stable for the sNormDist and sSumDist semantics. The average precision @ $N_{kf}$, shown in Figure 7, indicates the relevancy of the top percentile (max(1, k/f)) of search results with respect to the search size (k).

## 3.2. Vote-based Classification

To improve classification prediction, the scores of semantics, each classified by ω1, were input into the voting process. Two different voting functions were measured: $dG(\cdot, \cdot)$ and $dL(\cdot, \cdot)$. Both functions use spherical quadtrees and the rhombic partitioning technique (section 1.5) to partition spherical surfaces. Quad-trees were constructed with the following depths: 0 (no child nodes, only root nodes), 1, and 2. Deeper trees required more time to align than more shallow trees.

Only proteins that were classified differently by a pair of semantics were passed on to the voting process. Results from the voting process (number of proteins classified correctly vs. incorrectly) were merged with the classification results from the paired semantics to generate the final classification score. Both voting functions were applied to the following semantic pair:

$$\left( sRad_{k=1}^{14/14}, sNormDist_{k=2}^{15/16} \right)$$

and the results are summarized in Figure 8.

The $dG(\cdot, \cdot)$ function produced the highest scoring classifications in combination with centroids from $\Psi bb$. Each of the voting algorithms made use of an internal distance function based on the average pairwise distance. $dG(\cdot, \cdot)$ applied the average pairwise distance function as its distancemetric. $dL(\cdot, \cdot)$ applied average pairwise distance to each ring segment to produce a vector of distances, one distance for each ring; distance between the ring vectors was then computed using $L^2$.

dL sampled the protein space using variable-width ring segments such that each ring contained roughly 10 percent of the protein density (reused sRad-based segmentation). For both the dG and dL functions, voting was executed with the following values of $\Delta\theta$ (in degrees): 3, 4, 5, 6, 8, 10, 12, 18.

## 4. Discussion

Building the protein set W1 presented several challenges. Many revisions of SCOP have been released, and in the latest revision one of the query ($Q_1$) proteins has been reclassified such that it belongs in the same superfamily as another, chains 1din-A and 1xyz-E are both currently classified as superfamily c.69.1. The result is that there really only 179 unique superfamilies. However, to maintain consistency with prior work, the same query set was reused without modification. A side-effect of that decision is that superfamily c.69.1 actually has 19 members in $W_1$ as opposed to all of the others, which have 10.

Efforts were made to identify which version of SCOP was used to produce 180 unique superfamilies, the query proteins, and 1800 non-overlapping $W_1$ proteins. No such version of SCOP was identified. In addition, the testing performed in DDPIn was executed against SCOP version 1.65, an older version of the database.

Of the 1799 proteins constituting $W_1$, each was chosen to avoid overlap with those from the query set. It's not clear if prior works applied the same selection filter. If they did not, then that presents at least two concerns: (a) their query set may have contained protein chains that were duplicated in the searchable set ($W_1$), and; (b) if there were duplicate proteins between the two sets, then that could account for the higher results observed in their experiments. It is very likely that the W1 sets in DDPIn and this experiment are significantly different.

Generating the feature objects presented other challenges. Each PDB protein file is read from disk and converted to DCRR format in memory. As part of that process, centroid filtering is performed such that only one of the centroid sets is transformed ($\Psi$all, $\Psi$bb or $\Psi$sc) at a time. Data transformation operations are applied to the centroids to generate the feature objects.

The transformation process can be very time consuming depending on the nature of the semantic. This work dealt with that problem by leveraging many CPU resources to split up the work. A consequence of that decision was the tedious and time consuming data replication overhead required to keep each of the compute nodes in sync and up to date with the latest code and script changes. However, considering the many CPU hours required for the experiment, there was little other choice if the experiment was to be finished within the given time constraints. Transformation processes generally did not require much memory – the limited resource was the CPU.

Feature objects from both the $Q_1$ and $W_1$ were stored as compressed text files. While such a format is arguably not the most compact, it does have advantages. First, because it is a text file, there are a wide range of shell-based operations that may be performed on the file very easily. Some examples are verification of contents, summarizing proteins contained in a feature object, filtering proteins from a feature object, and adding new protein features to an existing object file. In addition, because compressed files have a smaller on-disk footprint, fewer I/Os are required when loading the database into memory, at the expense of additional CPU overhead to decompress the data. A smaller footprint is also more friendly to both hardware block level and OS file system level caching mechanisms.

What follows is an example of how this was useful to this experiment. Several database object files were corrupted by a bug in a semantic function, and as a result were crashing search operations. Only a small part of each database object file was actually corrupted (less than 3 percent). Due to the flexible nature of the format, it was trivial to determine which proteins had been compromised, strip them from the database object, regenerate smaller database object files for only those proteins, and then merge the newly generated objects back into the main database object file. This was very advantageous due to the many hours required to rebuild several complete searchable database object files from scratch.

Database objects were initially loaded into M-tree structures to prepare for search operations. The M-tree libraries that were integrated with the project presented their own challenges. First, there was *very* little documentation supplied with the libraries, leaving much room for guesswork. Source code for the library is not publicly available. Second, memory required by the M-tree structure started to become a limiting factor in terms of which compute nodes would be able to load the full database into memory. Finally, the M-tree structure proved to be unstable in several aspects, generating stack overflows and null pointer exceptions for certain semantic function configurations. Both the old and new releases of the library suffered from stability issues.

Problems with the M-tree led the research to consider the onion-tree structure. The onion-tree presented several advantages, the first of which was that the complete source code was available online. In addition, the algorithm was concise and easy to integrate with the rest of the experiment. Onion-tree was designed to be used as an in-memory database and as such presented a very compact in-memory representation that allowed all object databases to be loaded on all available compute nodes. Finally, the onion-tree structure had been optimized to reduce the number of distance calculations required for kNN and range based queries, the result of which was faster observable search times, when compared to those of M-tree. As such, much of the testing performed in this experiment made use of the onion-tree, though some of the early results generated from M-tree were kept.

Significant effort was put into continuously refactoring both the search and feature extraction processes to scale properly across a variety of compute node configurations. Several off-the-shelf Java parallel programming libraries were investigated as candidates to manage the parallelism within the algorithms. None of those evaluated met this work's requirements with respect to all of the following: (a) ease of integration to the existing code base; (b) run-time overhead, and; (c) configurability and extendability. Consequently, a custom parallel programming library was written to support the experiment. The interface is fairly high level and could be adapted, without much effort, to wrap a third party threading library, should one surface that meets the needs of this work.

Classification results were consistently better for searches using $\Psi$bb models than the others. Initial thinking was that the richer data set provided by $\Psi$all and $\Psi$sc models would also support consistently high-scoring classifications. This proved not to be the case.

The classification results are somewhat disappointing. Based on results of prior work, in particular DDPIn, the expectations were high for this experiment — especially for the semantic that replicates the sRad algorithm from DDPIn. In fact, the sRad semantic was expected to act as a stable baseline. The performance difference of sRad between DDPIn and this experiment is remarkable. DDPIn claims to have achieved a classification success rate above 90 percent with their sRad implementation, while this experiment's implementation of sRad peaked at just under 74 percent using the same classifier.

One possible reason for this difference is the implementation details of the segmentation algorithm. While the underlying details of DDPIn's sRad implementation are unknown, this work's implementation grouped "leftover" centroids (remainders of $|\Psi x|$/segments) such that they were distributed in sequential fashion starting with the inner-most ring segments and moving outward. An alternative (untested) implementation might attempt to distribute them more evenly over the available ring segments to avoid a consistently higher density of centroids in the inner-most rings. DDPIn's weighting scheme was also slightly different, squaring the weight vector components as part of the weighted $L^2$ calculation.

A more obvious potential cause for misclassification stems from the fact that the searchable protein set, $W_1$, was probably constructed much differently in DDPIn than in this experiment. As stated earlier, this work picked proteins at random and avoided duplicating proteins in $Q_1$. Minimal additional filtering was applied to the selection process. DDPIn's selection process is unclear, however, it is likely that more effort was made to obtain proteins in the $W_1$ set that are more representative of their respective superfamily. A selection process focused on higher-quality $W_1$ proteins could significantly change the outcome of the classification process. It is also possible that in DDPIn's experiment there were overlaps between the protein chains in $Q_1$ and $W_1$.

The original intent was to measure the effectiveness of the classifiers using precision and recall metrics calculated from the kNN search results. However, it quickly became apparent that some changes to the metric calculation would be appropriate. For example, with a kNN search, the there is no difference between the total number of objects retrieved (divisor of precision) and the total number of relevant objects that could have been retrieved (divisor of recall) - they're both equal to k. Since precision and recall reduce to the same value, it was decided that only precision would be reported.

Precision can be computed with respect to some top percentile of the returned objects, called precision@N, where N has a value greater than zero and less than the total number of objects retrieved. The graphs in Figure 7 show that while sRad and sNormDist show similar precision at low k values, sNormDist reports a higher precision for more top-level entries than sRad. This is an important behavioral difference between the two semantics: A search operation using sRad is more likely to retrieve a correct single match, but the same search using sNormDist is more likely to retrieve several correct matches.

There is a small group of proteins for which all combinations and configurations of semantic and voting functions failed to properly classify. It is possible that these proteins may not be representative of their respective superfamilies. Further investigation is required. The list of unclassifiable proteins appears in Table 5 and includes the centroid count for each protein along with its SCOP classification.

Voting did significantly improve classification results, bringing the top score from 73.889 percent to 80.000 percent. An important difference between DDPIn's voting mechanism and that of this experiment is that DDPIn used a Smith-Waterman DP procedure based on protein sequence while this work applied one of two purely structural distance functions.

The $dL(\cdot, \cdot)$ function is much more flexible than its $d_G$ counterpart. For example, ring segment size is configurable and can be set to use fixed widths, or percentiles based on the protein density. Unfortunately, due to time constraints, a very limited set of configurations was tested with the $d_L$ function. It is very possible that better classification scores are attainable by adjusting the parameters of the $d_L$ function.

As it currently stands, the tested configuration of $d_L$ performed worse than $d_G$. This may be due to the fact the the tested configurations of $d_L$ used the same radial segmentation algorithm as sRad. Given that fact, along with the knowledge that $d_L$ was voting on proteins that sRad and sNormDist had disagreed upon, it is quite possible that $d_L$ generated biased results. Other configurations of $d_L$ could be much less biased.

# 5. Future Work

Several opportunities exist for future work. The existing weighted $L^2$ distance function applies a non-normalized weight vector instead of one that has been normalized to unit vector form, which could cause an unintended distortion of the original feature vector values. Future work could improve upon this by simply normalizing the weight vector before applying the weighted function.

Second, there is a need for a more modern, standardized working protein set for both protein query ($Q_1$) and database ($W_1$) objects. The data sets used in this experiment are based on those of prior experiments which were performed long before the latest revision of the SCOP database. In addition, more care could be taken when selecting proteins for this data set to ensure that the selected proteins are indeed truly representative of their respective superfamilies.

New semantic functions could also be developed for use in the feature extraction process. For example, an sDistDist semantic that would compare the spherical distributions between two sets of centroid shells, perhaps based on the average minimum pairwise distance function.

3D Zernike (3DZ) descriptors have been shown to be useful for representing objects in spherical space [4] [5]. Future research would do well to apply them to the structural indexing process described in this research. As something to keep in mind, it is quite costly to build the database and query objects with the comparatively simpler algorithms that were tested in this experiment. An implementation of a feature extraction process that uses 3DZ would most likely be very heavy computationally, and so it would be important to have access to a large number of computing nodes.

Given that additional CPU nodes would speed up this research, future work should also investigate distributed application platforms for simplifying the process of leveraging many compute nodes. Map-reduce algorithms are a current trend in large scale parallel computing and several frameworks have been successfully developed [27] [28] [29]. It seems quite possible that the algorithms presented in this work could be ported to execute in a map-reduce environment. Ray-tile intersection seems a likely candidate for optimization in a map-reduce framework running on graphics processing units (GPUs).

Voting could be enhanced with additional, pluggable voting mechanisms. Ideally the additional voting algorithms would retain the structure-based essence of this work. For example, Matt is a structurally-based DP algorithm (vs. sequentially-based DP) [30]. Matt is able to perform multiple alignments, and uses local protein structure geometry in its DP calculations. The implementation of Matt is freely available [31].

In addition, the set of proteins applied in the voting process could be adjusted. The current implementation only votes on the proteins in the nearest neighborhood of each independent classification process. Another approach would be to additionally include in the voting process all other proteins belonging to the superfamilies of those proteins that appeared in the same neighborhood.

The process of assigning spherical partition values could also be enhanced. Instead of assigning partitions simple count values (like the current implementation), a deeper spherical quad-tree could be created to yield very small partitions. These partitions could be treated as pixels on a spherical screen. When centroids are projected, they could be drawn as Gaussian splats on the screen, instead of being projected as single-point rays. Overlapping spats would cause a pixel to have a higher, darker value. This approach should give better detail as to the overlap and proximity of neighboring centroids, however the number of partitions is expected to be greater and therefore the computational time is expected to significantly increase.

As part of an effort to scale this research to support larger data set sizes, future research should consider on-disk, or hybrid on-disk/in-memory MAM-based index structures. The size of the data set in this experiment is somewhat limited in scope; a larger data set would require more memory. Although the available memory in computers today significantly outpaces that which was available in prior computer generations, there still exists a tangible limit, especially for research institutions on a budget. A comprehensive data set encompassing all known proteins would likely not fit into memory and would need to be accessed on-disk, which would introduce performance issues due to the higher latency required for disk-based I/O. As an example of an MAM-based index with such support, a recently published cousin of M-trees, the Nested Approximate eQuivalence class (NAQ)-tree [32] seems to offer a significant performance improvement for disk-based MAMs. Another alternative would be to maintain the in-memory design and instead fragment the database, distributing it over multiple compute nodes, and modifying the search/scoring algorithms to take this fragmentation into account.

Finally, a Web-based interface to this search engine would be ideal for sharing this work with other researchers. An HypertextMarkup Language (HTML), web-based interface would allow single-protein queries, given a PDB input file. Candidates for web automation interfaces include a Web Services-based interface as well as a Representational State Transfer (REST)-base interface.

A Web-based Application Programming Interface (API) would offer, at a minimum, the following services:

- compare(A : protein,B : protein) : score

- neighbors(A : protein,N : howMany) : (P : List{pdbId}, S : List{score})

# 6. Conclusion

This experiment breaks new ground in several ways, the first of which is the use of the DCRR model for protein structure indexing. DCRR models were also used in the structural alignment of proteins during the voting process. In addition, this work develops and applies the concept of protein features generated fromcentroid spherical distance distributions for the purpose of structural protein indexing. Furthermore, spherical quad trees are built, not using trixels, but instead using a polyhedron of rhombic faces, which results in more uniform partitioning of a spherical surface. This work could be easily generalized for application to other centroid based representations for different types of molecules, for example carbohydrates.

## 6.1. Acronyms

| | |
|---|---|
| **3DZ** | 3D Zernike |
| **AFP** | aligned fragment pair |
| **API** | Application Programming Interface |
| **CDF** | cumulative distribution function |
| **CE** | Combinatorial Extension |
| **DCRR** | Double Centroid Reduced Representation |
| **DDPIn** | Distance and Density based Protein Indexing |
| **DP** | Dynamic Programming |
| **GPU** | graphics processing unit |
| **HTML** | HypertextMarkup Language |
| **JVM** | Java virtual machine |
| **kNN** | k-nearest neighbor |
| **MAM** | Metric access method |
| **MDI** | multi-dimensional indexing |
| **M-tree** | metric tree |
| **NAQ** | Nested Approximate eQuivalence class |
| **PDB** | Protein Data Bank |
| **REST** | Representational State Transfer |
| **RMSD** | Root Mean Square Deviation |
| **SCOP** | Structural Classification Of Proteins |
| **SSAP** | Sequential Structure Alignment Program |

## 6.2. Glossary

$\Psi$ symbol for centroid. $\Psi c$ is the symbol used to represent a class of centroids for one or more proteins, where c may indicate all centroid classes ($\Psi all$), the class of backbone centroids ($\Psi bb$) or the class of side-chain centroids ($\Psi sc$). $\Psi x$ is written to represent a singular centroid of an arbitrary class, page 1.

$\mathbb{H}$ Hamiltonian, or quaternion space. Quaternions are represented using a single real component and three imaginary components $q = w+ix+jy+kz$, where i, j, k are imaginary. The conjugate of a quaternion is found by negating the imaginary components (but not the real component). In this work quaternions are applied to simplify the rotation of protein structures around arbitrary axes, page 8.

$\omega$ classification function, page 11.

$\Re^n$ n-dimensional Euclidean space, page 3.

$d_G$, $d_L$ protein alignment algorithms that, for two or more proteins, project centroids of each protein to a partitioned spherical surface. Centroids are rotated systematically and the distance between partition sets is computed at each rotation. The best alignment occurs at the minimum distance between two partitioned surfaces. dG uses a single, global partition set for each protein. dL uses concentric partition sets, page 13.

$L^p$        vector space in which p-norms of such vectors are referred to as Lp norms, or Lp distance. L2 is Euclidean distance, page 4.

$S^2$        unit sphere in $\Re 3$ with a radius of 1, page 10.

$sRad_{k=z}^{x/y}$ Shorthand notation for a query configuration (most likely associated with some set of classification results) using features generated from a specific semantic. In this example, sRad is the semantic, x represents the number of feature dimensions that were input into the structural distance function (all dimensions are only used for some configurations), y indicates the overall number of feature dimensions and z is the size of the kNN search, page 14.

**k-nearest neighbor**     is a non-parametric technique for building (or identifying) a cluster of n data points around a sample point x, and is implemented by expanding a "window" or "cell" size around the sample point x until the window encompasses n data points. The related probability density function can be written as $p_n(x) = (k_n/n)/V_n$, where $V_n$ is the volume of the window [11], page 3.

**centroid**       Centroids are computed from the geometric average of a collection of atoms that constitute some significant aspect of an amino acid. For example, this work makes use of backbone centroids and side-chain centroids. For a given amino acid, the backbone centroid is computed from the N, O, C and C_ atoms of the acid's backbone. The sidechain centroid is computed from the molecules that make up the R-group of the amino acid, page 1.

**M-tree**        metric access method used to organize and search large data sets from a generic "metric space [15], page 5.

**spherical distance**    is the great circle distance measured between two points which lay on the surface of a sphere; it is the length of the shortest arc that may be drawn between the two points, page 9.

**Wasserstein metric**    also known as *earth mover's distance*, is the distance between two probability distributions in a metric space, page 4.

## 6.3   References

[1]  C. A. Orengo and W. R. Taylor, "SSAP: Sequential structure alignment program for protein structure comparison," in *Computer Methods for Macromolecular Sequence Analysis* (R. F. Doolittle, ed.), vol. 266 of *Methods in Enzymology*, pp. 617 − 635, Academic Press, 1996.

[2]  I. N. Shindyalov and P. E. Bourne, "Protein structure alignment by incremental combinatorial extension (CE) of the optimal path," *Protein Engineering*, vol. 11, no. 9, pp. 739–747, 1998.

[3]  V. M. Reyes and V. N. Sheth, "Visualization of protein 3D structures in 'Double-Centroid' Reduced Representation: Application to ligand binding site modeling and screening," *Handbook of Research in Computational and Systems Biology: Interdisciplinary Approaches*, 2010. In press.

[4]  L. Mak, S. Grandison, and R. J. Morris, "An extension of spherical harmonics to region-based rotationally invariant descriptors for molecular shape description and comparison," *Journal of Molecular Graphics and Modelling*, vol. 26, no. 7, pp. 1035 − 1045, 2008.

[5]  M. Novotni and R. Klein, "Shape retrieval using 3D zernike descriptors," *Computer-Aided Design*, vol. 36, no. 11, pp. 1047 − 1062, 2004. Solid Modeling Theory and Applications.

[6]  P. Papadakis, I. Pratikakis, S. Perantonis, and T. Theoharis, "Efficient 3D shape matching and retrieval using a concrete radialized spherical projection representation," *Pattern Recogn.*, vol. 40, pp. 2437–2452, September 2007.

[7] M. Chaouch and A. Verroust-Blondet, "Alignment of 3D models," *Graph. Models*, vol. 71, pp. 63−76, March 2009.

[8] H. Rangwala and G. Karypis, "fRMSDPred: Predicting local RMSD between structural fragments using sequence information," *Proteins*, vol. 72, no. 3, pp. 1005 − 1018, 2008.

[9] M. Mechelke and M. Habeck, "Robust probabilistic superposition and comparison of protein structures," *BMC Bioinformatics*, vol. 11, no. 1, p. 363, 2010.

[10] D. Hoksza, "DDPIn: distance and density based protein indexing," in *Proceedings of the 6th Annual IEEE conference on Computational Intelligence in Bioinformatics and Computational Biology*, CIBCB'09, (Piscataway, NJ, USA), pp. 263−270, IEEE Press, 2009.

[11] R. O. Duda, P. E. Hart, and D. G. Stork, *Pattern Classification*. New York, NY: John Wiley and Sons, Inc., 2nd ed., 2001.

[12] X. Rong Li and Z. Duan, "Comprehensive evaluation of decision performance," in *Information Fusion, 2008 11th International Conference on*, pp. 1 −8, July 2008.

[13] J. Wang, X. Wang, D. Shasha, and K. Zhang, "MetricMap: an embedding technique for processing distance-based queries in metric spaces," *Systems, Man, and Cybernetics, Part B: Cybernetics, IEEE Transactions on*, vol. 35, no. 5, pp. 973 −987, 2005.

[14] J. Zhou, J. Sander, Z. Cai, L.Wang, and G. Lin, "Finding the nearest neighbors in biological databases using less distance computations," *Computational Biology and Bioinformatics, IEEE/ACM Transactions on*, vol. 7, no. 4, pp. 669 −680, 2010.

[15] P. Ciaccia, M. Patella, and P. Zezula, "M-tree: An efficient access method for similarity search in metric spaces," in *Proceedings of the 23rd International Conference on Very Large*

*Data Bases*, VLDB '97, (San Francisco, CA, USA), pp. 426–435, Morgan Kaufmann Publishers Inc., 1997.

[16]   T. Skopal, J. Pokorn`y, M. Kr`atk`y, and V. Sn`aˇsel, "Revisiting M-Tree Building Principles," in *Advances in Databases and Information Systems*, vol. 2798 of *Lecture Notes in Computer Science*, pp. 148–162, Springer Berlin / Heidelberg, 2003.

[17]  P. Ciaccia, F. Rabitti, M. Patella, and P. Zezula, "The M-tree Project." http://www-db.deis.unibo.it/Mtree/, 2008.

[18]   J. Sedmidubsk´y and V. Dohnal, "M-tree." http:// lsd.fi.muni.cz/trac/mtree/, 2010.

[19]   C. C. M. Carlo, I. R. V. Pola, R. R. Ciferri, A. J. M. Traina, C. T. Jr, and C. D. de Aguiar Ciferri, "Slicing the metric space to provide quick indexing of complex data in the main memory," *Information Systems*, vol. 36, no. 1, pp. 79 − 98, 2011. Selected Papers from the 13th East-European Conference on Advances in Databases and Information Systems (ADBIS 2009).

[20]   C. C. M. C. Ives Renˆe Venturini Pola, "Onion-tree." http://gbd.dc.ufscar.br/download/Onion-tree, 2009.

[21]   V. M. Reyes, "Representation of protein 3D structures in spherical (ρ,φ,θ) coordinates and two of its potential applications." Unpublished manuscript, 2010.

[22]  E. W. Weisstein, "Kulikowski's Theorem." http://mathworld.wolfram.com/KulikowskisTheorem.html, 2010. From MathWorld−AWolframWeb Resource.

[23]   G. Fekete, "Rendering and managing spherical data with sphere quadtrees," in *Proceedings of the 1st conference on Visualization '90*, VIS '90, (Los Alamitos, CA, USA), pp. 176–186, IEEE Computer Society Press, 1990.

[24] A. G. Murzin, S. E. Brenner, T. Hubbard, and C. Chothia, "SCOP: A structural classification of proteins database for the investigation of sequences and structures," *Journal of Molecular Biology*, vol. 247, no. 4, pp. 536–540, 1995.

[25] O. Ça̧moglu, T. Kahveci, and A. K. Singh, "Towards index-based similarity search for protein structure databases," in *Proceedings of the IEEE Computer Society Conference on Bioinformatics*, CSB '03, (Washington, DC, USA), pp. 148–, IEEE Computer Society, 2003.

[26] H. M. Berman, J. Westbrook, Z. Feng, G. Gilliland, T. N. Bhat, H. Weissig, I. N. Shindyalov, and P. E. Bourne, "The Protein Data Bank," *Nucleic Acids Research*, vol. 28, pp. 235–242, 2000.

[27] T. Apache Software Foundation, "Hadoop." http: //hadoop.apache.org, 2011.

[28] B. He, W. Fang, Q. Luo, N. K. Govindaraju, and T. Wang, "Mars: A MapReduce Framework on Graphics Processors," in *PACT'08: PROCEEDINGS OF THE SEVENTEENTH INTERNATIONAL CONFERENCE ON PARALLEL ARCHITECTURES AND COMPILATION TECHNIQUES*, pp. 260–269, 2008. 17th International Conference on Parallel Architectures and Compilation Techniques, Toronto, CANADA, OCT 25-29, 2008.

[29] M. M. Rafique, B. Rose, A. R. Butt, and D. S. Nikolopoulos, "CellMR: A Framework for Supporting MapReduce on Asymmetric Cell-Based Clusters," in *2009 IEEE INTERNATIONAL SYMPOSIUM ON PARALLEL & DISTRIBUTED PROCESSING, VOLS 1-5*, International Parallel and Distributed Processing Symposium (IPDPS), pp. 883–894, 2009. 23rd IEEE International Parallel and Distributed Processing Symposium, Rome, ITALY, MAY 23-29, 2009.

[30] M. Menke, B. Berger, and L. Cowen, "Matt: Local flexibility aids protein multiple structure alignment," *PLoS Comput Biol*, vol. 4, p. e10, 01 2008.

[31]   M. Menke, B. Berger, and L. Cowen, "Matt: Local flexibility aids protein multiple structure alignment." http://groups.csail.mit.edu/cb/matt/, 2007. Preprint.

[32]   M. Zhang and R. Alhajj, "Effectiveness of NAQtree as index structure for similarity search in highdimensional metric space," *Knowledge and Information Systems*, vol. 22, pp. 1–26, 2010.

## 7.1 FIGURES and Legends:

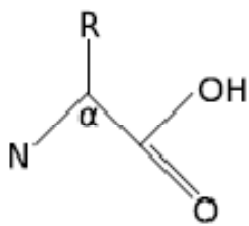

**Figure 1**: Chemical structure of an amino acid. The _ represents the C_ atom, the R symbol represents the R-group, or side-chain structure attached to the amino acid. When the R group is present, C_ is in the R position.

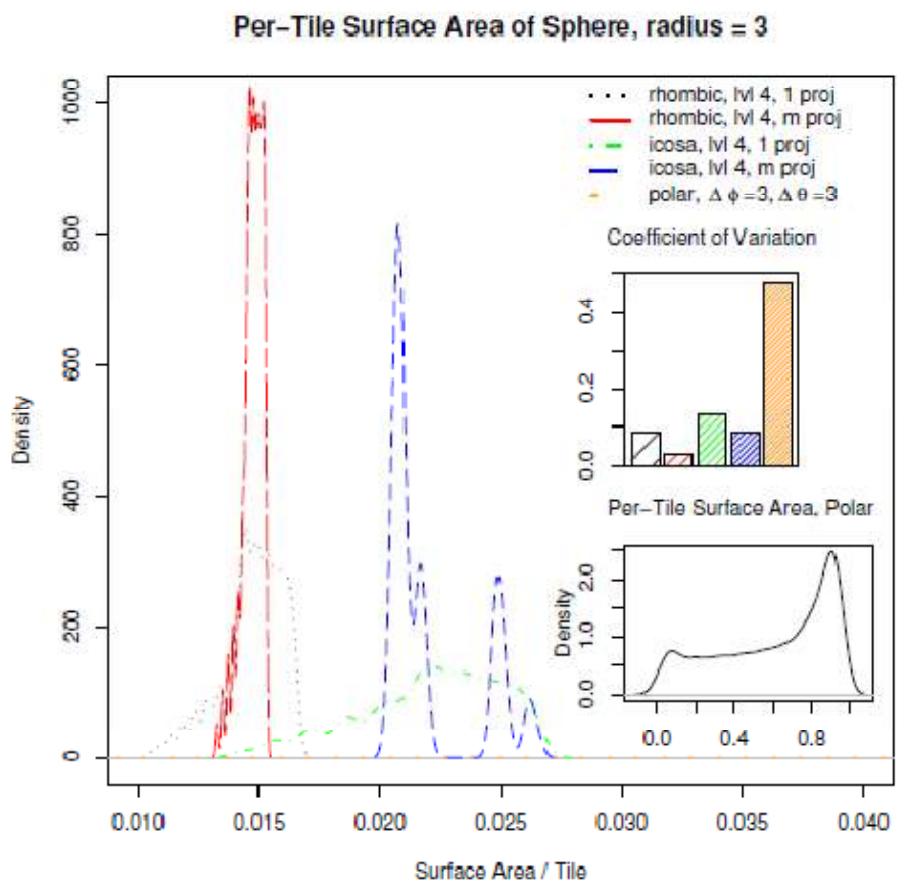

**Figure 2:** Observe that the rhombic mesh using multiple iterations of bisection/projection has the tallest, densest peak (red, longest dashes), yielding the most uniform tiles. Meshes generated from icosahedrons provided less uniform tiles. Polar binning generates least uniform tiles (flat distribution, sparse orange, dotted line along the bottom), A : [0.0167,0.940]. The number of tiles for the rhombic, triangle and polar meshes are 7680, 5120 and 7320 respectively. For each polyhedron-derived mesh, iterations of bisection/ projection provided more uniform tile sizes. The coefficient of variation cv = _ μ is lowest (best, 0.028) for the rhombic mesh, level 4, multi-bisection/projection algorithm; it's highest (worst, 0.481) for polar binning mesh.

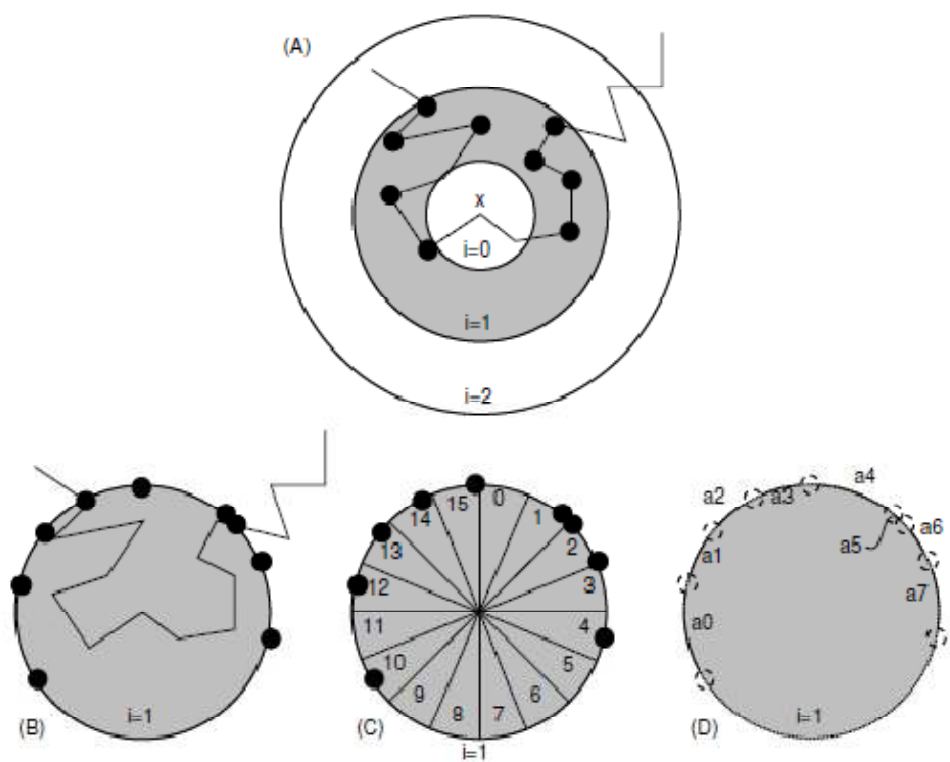

**Figure 3:** 2D example of feature extraction of value Px i , where x indicates       x and i = 1 identifies the current sphere (or circle in this case) under analysis. Initial protein centroids are partitioned according to concentric spheres (A), then projected to the surface of the nearest outer sphere (B). Different feature extraction methods analyze distribution of centroids in different

ways: binning (C), and distance (D). (D) shows a subset of the arcs in the set of spherical distances used for some semantics.

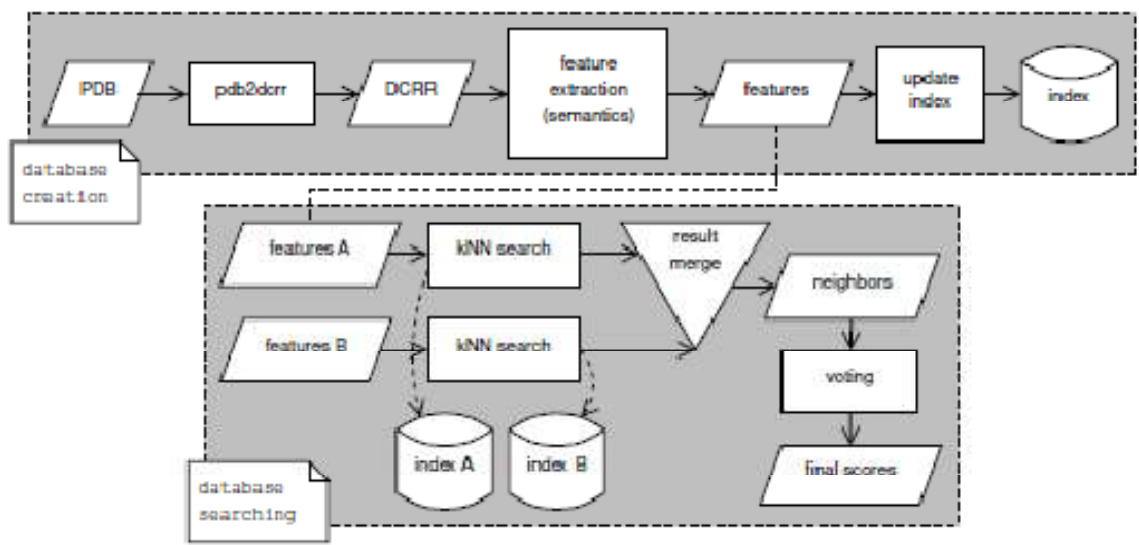

**Figure 4:** Database construction: PDB files converted to DCRR format and features are extracted, then loaded into an in-memory index. Queries: PDB-based queries are converted into DCRR format, and extracted features are compared to the indices to determine best matches (kNN). Matches are aggregated, then filtered by a voting process that picks the best results.

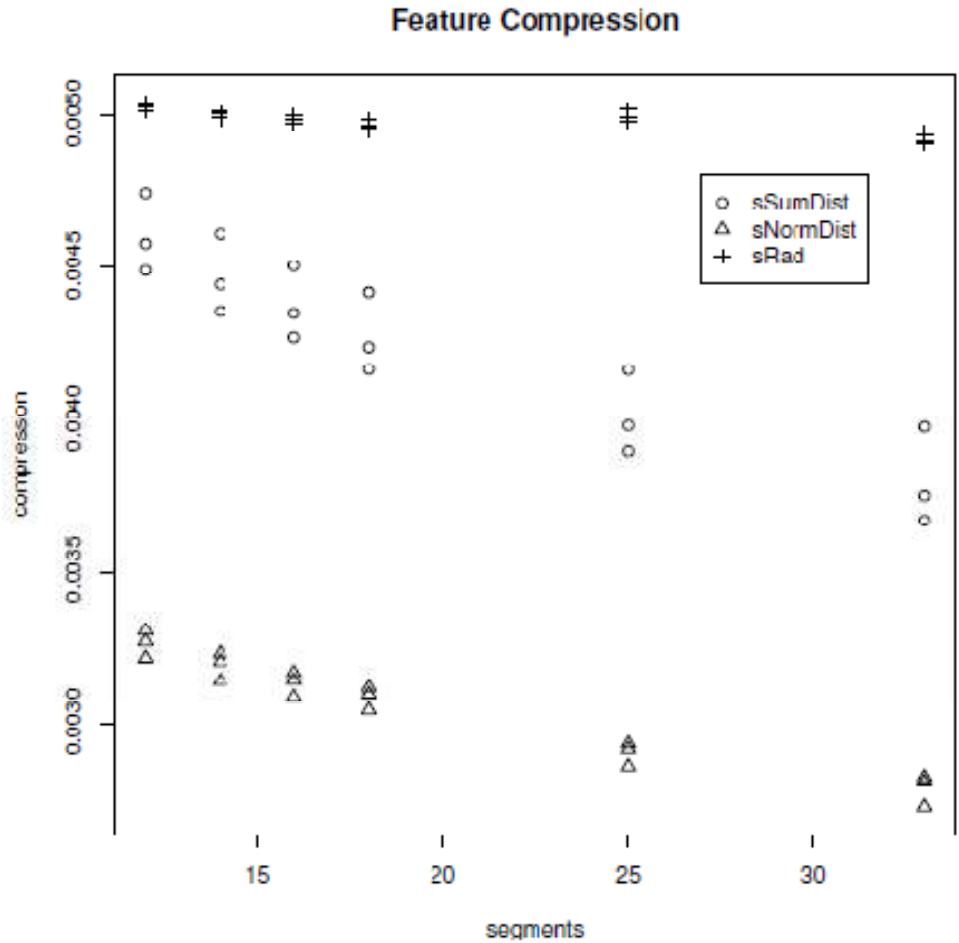

**Figure 5:** Feature object compression factors across all centroids, semantics and segment sizes. A higher compression factor indicates a less compressible feature object set. sRad semantics consistently produce the least compressible feature object sets.      all centroid sets show consistently larger compression factors than  bb or   sc sets.

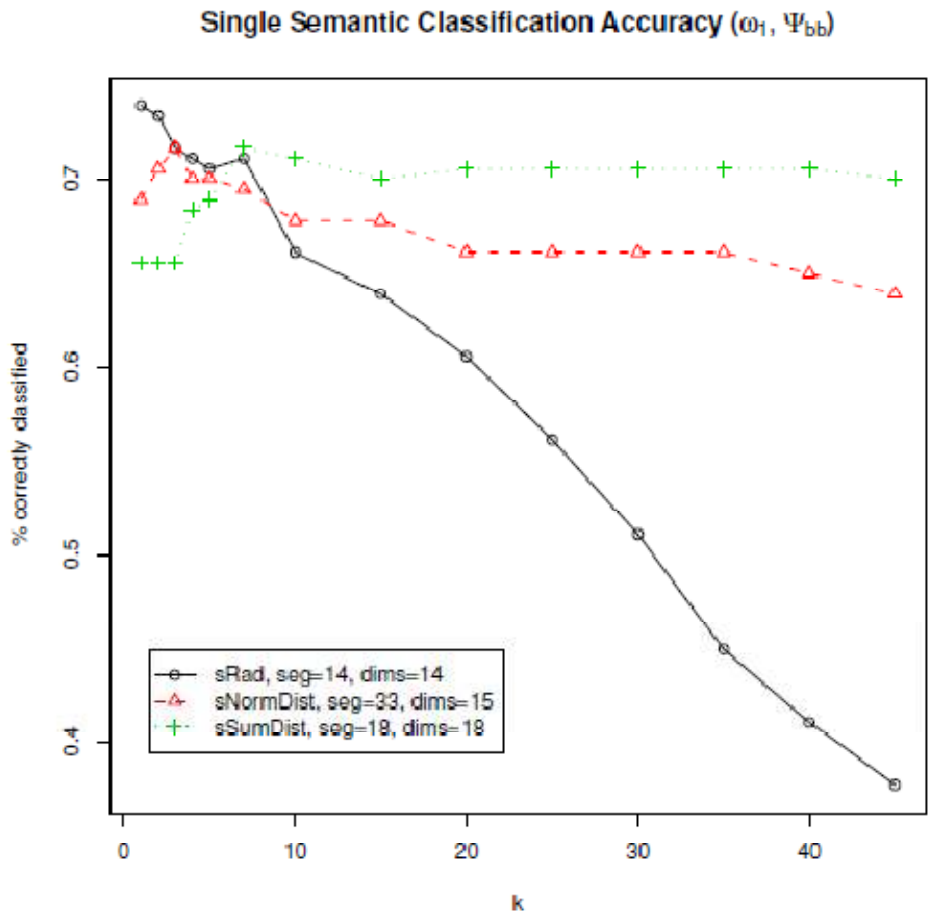

**Figure 6:** sRad provides the best classification at very small k values. However, sRad classification rapidly deteriorates as k increases whereas sNormDist and sSumDist are much more stable.

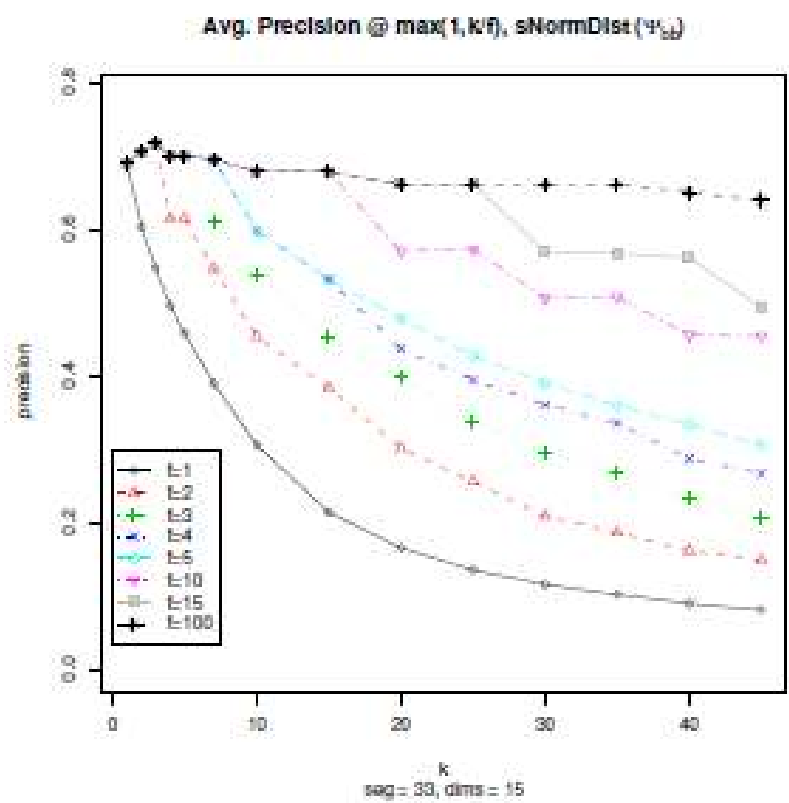

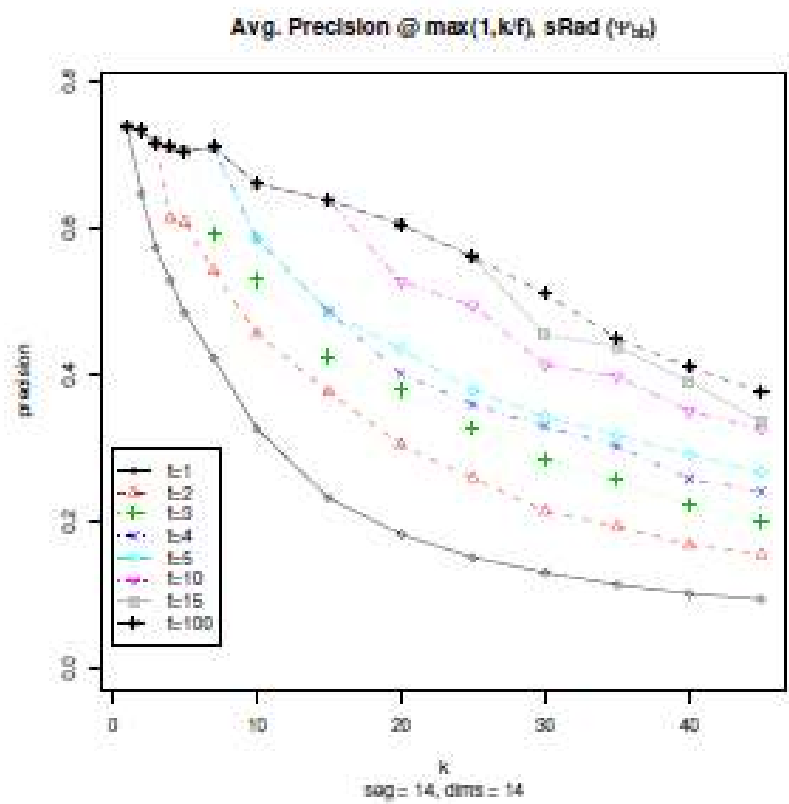

**Figure 7:** Precision metrics for the sNormDist and sRad semantics.Both semantics give rise to more precise classifications at lower k values. sRad gives the best results at k = 1, sNormDist gives more high scoring results as k continues to increase.

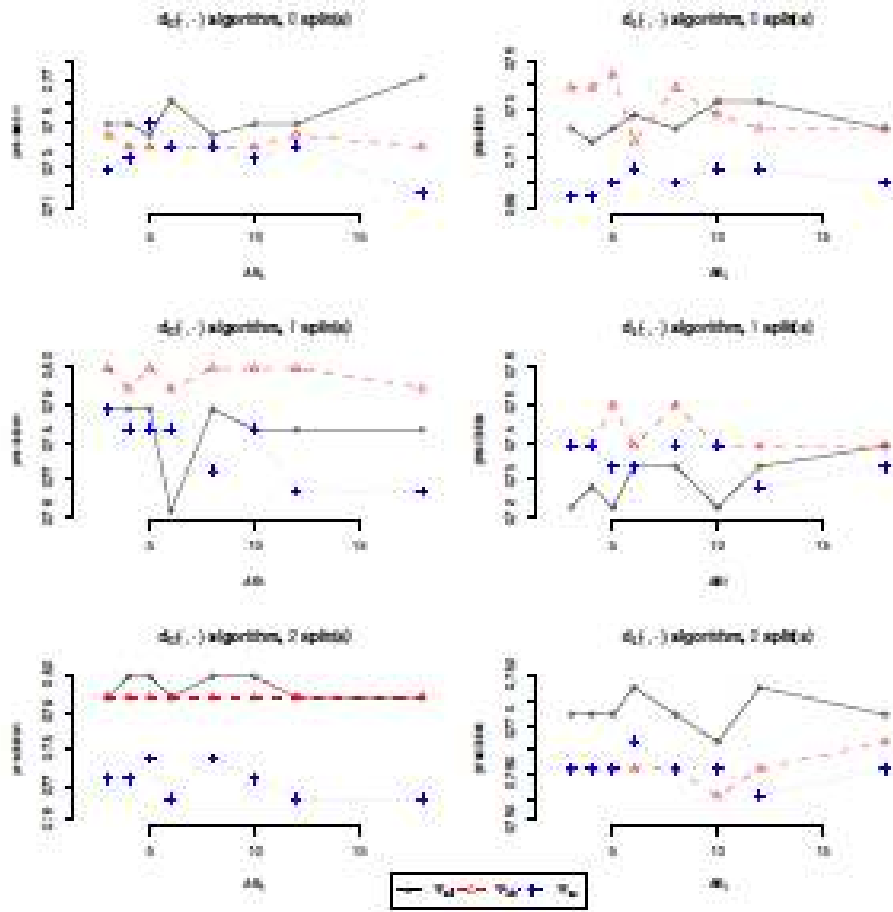

**Figure 8:** dG performs well using quad-trees of depths 1 and 2, though for different centroid classes ( bb and all). dL underperforms when compared to dG, but shows some improvement using deeper quad-trees.

## 7.2    **TABLES and Legends:**

| semantic | distance func. | segments (dims) | |
|---|---|---|---|
| sRad  sSumDist  sNormDist | $L^2_{weighted}$ | 12 | (9, **10**, 11, 12) |
| | | 14 | (11, 12, 13, 14) |
| | | 16 | (13, 14, 15, 16) |
| | | 18 | (15, 16, 17, 18) |
| | | 33 | (11, 13, 15, 17, 19, 21, 23, 25, 27) |

**Table 1:** All of the semantic configurations used the same L2 weighted distance function as well as the same set of dims/segments parameters.

| CPU spec | # nodes | memory |
|---|---|---|
| Amazon EC2 (high CPU, x-large) | 4 | 7GB |
| dual-core AMD, 2GHz | 1 | 4GB |
| 2x4-core Xeon, 2.8GHz | 1 | 4GB |
| 2x4-core Xeon, 3.2GHz | 1 | 8GB |
| 2x12-core AMD, 2.1GHz | 1 | 16GB |
| 4x4-core AMD, 2.3GHz | 1 | 32GB |

**Table 2:** Many compute nodes are required for feature object builds as well as database searches. High CPU, x-large EC2 servers are configured with 20 EC2 units distributed over eight cores.

| semantic | $\Psi$ | segments | time |
|----------|--------|----------|------|
| $sRad$ | all | 17 | 10m |
| $sRad$ | bb/sc | 17 | 5m |
| $sRad$ | all | 33 | 32m |
| $sRad$ | bb/sc | 33 | 10m |
| $sNormDist$ | all | 18 | 51h |
| $sNormDist$ | bb/sc | 18 | 8.5h |
| $sSumDist$ | all | 18 | 36h |
| $sSumDist$ | bb/sc | 18 | 9.25h |

**Table 3:** Sample feature object build times for W1 proteins. Features generated from more complex semantics take longer to build.

| Ψ | semantic | segments | k | dims | score |
|---|----------|----------|---|------|-------|
| $\Psi_{bb}$ | $sRad$ | 14 | 1 | 14 | 0.7388889 |
| $\Psi_{bb}$ | $sRad$ | 16 | 1 | 16 | 0.7388889 |
| $\Psi_{bb}$ | $sRad$ | 18 | 1 | 17 | 0.7388889 |
| $\Psi_{bb}$ | $sRad$ | 12 | 1 | 12 | 0.7166667 |
| $\Psi_{bb}$ | $sNormDist$ | 33 | 3 | 15 | 0.7166667 |
| $\Psi_{bb}$ | $sSumDist$ | 18 | 7 | 18 | 0.7166667 |

**Table 4:** Top results of single semantic searches, classified by !1. sRad consistently scores better than sNormDist and sDist.    bb was the highest scoring centroid data set.

| PDB Chain | SCOP sf | $|\Psi_{all}|$ | $|\Psi_{bb}|$ | $|\Psi_{sc}|$ |
|-----------|---------|------------|-----------|-----------|
| 1c2n-A | a.3.1.1 | 4640 | 2320 | 2320 |
| 1cx1-A | b.18.1.14 | 6732 | 3366 | 3366 |
| 1kc6-B | c.52.1.19 | 487 | 247 | 240 |
| 1kpg-A | c.66.1.18 | 533 | 276 | 257 |
| 1ret-A | a.4.1.2 | 1462 | 731 | 731 |
| 1thj-C | b.81.1.5 | 410 | 213 | 197 |
| 1wiu-A | b.1.1.4 | 5580 | 2790 | 2790 |
| 2bby-A | a.4.5.15 | 4140 | 2070 | 2070 |
| 7cel-A | b.29.1.10 | 817 | 433 | 384 |

**Table 5:** Consistently misclassified proteins. No combination of search and/or voting parameters properly classified these. Notably absent from the list are protein chains in the d domain (_ + _).

**Table 6:** *Reference set of PDB IDs used in prior work [25], reused here to facilitate comparison with prior methods.*

| Query ids | Queries | | | | | | | | | |
|-----------|---------|---------|---------|---------|---------|---------|---------|---------|---------|---------|
| 1-10 | 1lh1 | 1c2n | 1ret | 2bby | 1wjc-A | 2spz-A | 2lef-A | 1b67-A | 1rcp-B | 1i4z-B |
| 11-20 | 1fr0-A | 1dps-I | 2cyk | 1unk-B | 1adr | 1tn4 | 1mj2-B | 1fk1-A | 1ihf-B | 1b4f-B |
| 21-30 | 1qck-A | 2ygs-A | 1hg4-B | 1kvw | 1i77-A | 1wiu | 1egj-A | 1ej8-A | 2msp-A | 1do6-A |
| 31-40 | 1g43-A | 1hu8-B | 1g1o-D | 3pcn-N | 1bqk | 1gmi-A | 1cov-3 | 1hdf-A | 1shs-A | 1slu-A |
| 41-50 | 1bd9-A | 1cx1-A | 1jh5-C | 1dmz-A | 7ce1 | 4hck | 2vub-A | 2pdz-A | 1i4k-S | 1pto-E |
| 51-60 | 1vqi | 1mjx-B | 1gus-C | 7i1b | 1ba7-B | 1inc | 3hvp | 1i7a-A | 1ief-B | 1acd |
| 61-70 | 2iza | 1dyw-A | 1bnu | 1bwu-P | 1hg8-A | 1thj-C | 1cax-F | 1hjg-A | 1qaw-B | 1a6x |
| 71-80 | 2f3g-B | 1kdf | 1a5l-B | 1f39-A | 1hg3-D | 1cw2-A | 1g4s-A | 1d3g-A | 1az2 | 2xyl |
| 81-90 | 1ez2-B | 1fxq-B | 1dxf-A | 1xic | 1ptd | 1dhr | 1dkd-C | 1b2s-D | 1tyf-L | 1c2y-P |
| 91-100 | 1f1j-B | 2chf | 1fln | 1xze | 1d0i-C | 1i7s-D | 1gn8-A | 1cd5-A | 1dts | 1jh8-A |
| 101-110 | 1sud | 5cev-A | 1qca | 1jf8-A | 1ypt-A | 1aiu | 1kc6-B | 1a5v | 1vfn | 1a2z-C |
| 111-120 | 8cpa | 1jlj-B | 2hpa-B | 1upu-D | 1bhq-2 | 1kpg-A | 1h6j-B | 1din | 1mas-A | 1drf |
| 121-130 | 1rk2-D | 1jdi-A | 7icd | 1qui | 4mt | 205l | 1au0 | 1bxi-B | 1rbg | 1e1s-A |
| 131-140 | 1ejr-A | 1qg7-B | 1azq-A | 3rhn | 1lfd-C | 1doy | 1c78-A | 1igd | 1gd3-A | 1e3v-A |
| 141-150 | 1ayz-B | 2emd | 1fkg | 1jc4-A | 1eyp-A | 2ci2-I | 1ec6-B | 1frk | 2nck-R | 1fj7-A |
| 151-160 | 1f9f-B | 1fe4-B | 1rcx-S | 1dch-C | 1xxb-F | 2cht-E | 1otf-D | 1icr-B | 4aig | 1bkl |
| 161-170 | 2hpr | 1b9l-A | 1i1d-A | 1hqz-8 | 1fil | 1byw-A | 1ga7-A | 1b5m | 1i5c-B | 1qmr-A |
| 171-180 | 1f7l-A | 1g3i-R | 1aha | 1prt-G | 1bnl-A | 1fzd-B | 1gu9-C | 1jya-A | 1is8-K | 1lep-E |

**Table 6:** Reference set of PDB IDs used in prior work [25], reused here to facilitate comparison with prior methods.